# Smoothed Particles Hydrodynamics numerical simulations of droplets walking on viscous vibrating liquid

Diego Molteni[1], Enrico Vitanza[2], Onofrio Rosario Battaglia[1*]

[1] Dipartimento di Fisica e Chimica, Università degli Studi di Palermo, Italy

[2] Dipartimento di Ingegneria Civile, Ambientale e Aerospaziale Università degli Studi di Palermo, Italy

**Abstract:**

We study the phenomenon of the "walking droplet", by means of numerical fluid dynamics simulations using the Smoothed Particle Hydrodynamics numerical method. This phenomenon occurs when a millimetric drop is released on the surface of an oil of the same composition, contained in a tank and subjected to vertical oscillations of frequency and amplitude very close to the Faraday instability threshold. At appropriate values of the parameters of the system under study, the oil droplet jumps permanently on the surface of the vibrating liquid forming a localized wave-particle system, reminding the behaviour of a wave particle quantum system as suggested by de Broglie. In our study, we made relevant simplifying assumptions, however we observe that the wave-drop coupling is surely obtained. Moreover, the droplet and the wave travel at nearly constant velocity, as observed in experiments. These facts suggest that the phenomenon may occur in many contexts and opens the possibility to study it in an extremely wide range of physical configurations.

[*] Corresponding author at: Dipartimento di Fisica e Chimica, Viale delle Scienze edificio 18, Palermo, Italy.

*E-mail address:* onofriorosario.battaglia@unipa.it (O.R. Battaglia).

## 1. Introduction

The fact that a solid particle can jump persistently on a vibrating rigid desk is trivial. Less obvious is the similar occurrence when the particle is an oil droplet falling on a vibrating viscous liquid. The phenomenon of "walking droplets" has been discovered rather recently by Y. Couder [1] and it is currently studied by various researchers. The phenomenon has prompted much attention due to the analogy that this classical fact exhibits with the quantum mechanical behaviour of particles, e.g. see Couder and Fort [2]. It has been shown [1] that in an appropriated range of parameters, a droplet, deposited on a vibrating bath of a liquid of the same composition, can bounce indefinitely and generates a localized wave strongly connected to the droplet, which can move at constant horizontal velocity. To obtain this peculiar behaviour the amplitude of the vibration must be very close to the threshold of the Faraday instability. The phenomenon of the Faraday instability is well known since 1831 [3]: a viscous liquid, subjected to vertical oscillations of amplitude greater than a well-defined critical value correlated to the frequency of the oscillations, the surface tension, and to the viscosity of the liquid, produces standing waves with a frequency half of the driving one. A simplified theoretical analysis predicts that, in a shallow liquid with height of the order of few millimeters, the wavelength is expected to be, approximately, inversely proportional to the forcing frequency [4]. The requirement of working slightly below the Faraday instability threshold is due to the fact that, in this case, the instability produces waves only in proximity of the droplet collision zone and therefore a localized wave is generated, while the remaining liquid oscillates as a whole, but no wave is produced. The particle-wave is affected by the boundary conditions and it may produce interference patterns and other typical quantum-like effects [5, 6]. It is a classical analogue of the de Broglie idea of the wave – particle dualism occurring on atomic and subatomic scales.

The interest on reproducing this phenomenon by a numerical simulation is relevant. Although some authors [7] propose tentative analytical equations for the motion of the individual drop and integrate them numerically taking into account the surface deformation, up to date, there are no truly fluid numerical simulation of this phenomenon. With affordable simulations, it is possible to investigate problems and configurations which are difficult to set up in laboratory. For example, it will be easily possible to investigate the role of an attractive force between two particles, like an elementary atom, or to make a linear oscillator and investigate the possibility of discrete energy levels, like the quantum oscillator. We will not discuss furthermore the quantum analogy, that, in our opinion, is far to be demonstrated (see also the arguments against the analogy by Andersen [8]).

In principle, the walking droplet problem is not very difficult to be studied with a Lagrangian numerical fluid dynamical method. The presence of waves deforming the liquid surface makes the Lagrangian codes better suited than the Eulerian ones. We studied this phenomenon using the Smoothed Particles Hydrodynamics (SPH) numerical method to simulate the motion of the oil in the vessel, while the droplet is simulated by a small group of points (or even a single point, i.e. with no internal structure) each interacting both with the other droplet points and with the discrete particles of the oil by *ad hoc* forces described below.

A more realistic simulation would require the treatment also of the air dynamics and not only of the oil. This approach had to manage also the large density difference between air and liquid. This kind of problems have been studied with the SPH methodology [9]. However, this full physical simulation, even if it is possible, has some practical drawbacks. To have a realistic simulation of the surface tension, the simulated droplet should have a spatial resolution such to guarantee a number of SPH particles of the order, say, of 500 particles [10], then the total number of particles (air, oil and droplet) is consequently rather large for a 2D (X-Z) case. Indeed, if we assume that a droplet has a 0.5 mm radius, and that it should be composed by 500 particles, we need a spatial resolution of about 0.00004 m, then, to reproduce a container of length 0.1 m and vertical height 0.01 m, including air, we need about 636,000 particles. Furthermore, the time step falls down since the *dt* due to the artificial viscosity scales as $h^2$ and for any kind of viscosity the diffusion condition requires $dt \leq \frac{h^2}{\sigma_{visc}}$ [11].

For this reason, we preferred to avoid the simulation of the air component, to mimic the interaction of the droplet with air and to take into account the surface tension by some ad hoc simple and successful physical models described below. The possibility to obtain significant results with a modest number of particles is, in our opinion, an added value, which can be exploited without special computational resources.

We set up a small 2D tank of viscous liquid, vibrating very close to the Faraday resonance frequency. A droplet is released on the liquid surface and, with some simplifying assumptions, we show that the basic phenomenon of the wave particle coupling is reproduced.

Many ingredients concur to produce this phenomenon. We tried to mimic the basic ones. Despite we made some strong approximations to make feasible the simulations, the fact that we reproduce the general experimental behaviour indicates that some parameters are more essential than other ones and that the phenomenon may be physically more ubiquitous.

We emphasize that the aim of this communication is basically to show that with a quite simplified model and a numerical simulation based on SPH method it is possible to study this interesting phenomenon.

The subject is treated as follows: in Section 2 we discuss the problem and its approximations; in Section 3 we briefly explain the SPH numerical method; in Section 4 we discuss the parameters of the systems that we simulate and show some of the main results. Finally, in Section 5, general conclusions are summarized.

## 2. The model and its approximations

The basic physical elements required to produce the phenomenon are: (1) - Vibration of the tank at a frequency very close to the Faraday instability acting as energy reservoir and pump. (2) - The specific amplitude of the oscillation, which can even produce accelerations larger than the gravitational one. (3) - The viscosity of the liquid, to stabilize the dynamics. (4) - The surface tension on the liquid and of the droplet to produce or enhance the bouncing force activated by the extra air pressure. (5) - The air cushion to produce the bouncing force.

The interaction of this localized wave with the droplet at a specific vibration frequency, amplitude, together with a specific size of the droplet, produces the coupled wave-particle synchronized motion. In our approach, we essentially studied two models of the droplet: one, a bit more realistic, in which the liquid droplet has a finite dimension, and it is made by a small group of points, held together by an appropriate small range force that produces a tension, and interacting with the liquid particles with another suitable force.

In a second model, extremely simple, the droplet is build up by a single point particle, i. e. a moving point, without any internal structure, interacting with the other liquid particles by an *ad hoc* force.

2.1 Interactions for the droplet made by points

2.1.1 The surface tension

With appropriate mathematical interpolation criteria for the boundaries, the SPH method can treat the surface tension of a liquid [12, 10]. Obviously, these approaches require a number of interpolating points sufficiently large to accurately calculate the curvature of the surface and all the related derivatives for the liquid under study. As mentioned in the previous section, in the phenomenon that we want to study the liquid droplet interacts with the liquid in the tank by the presence of the air, but we decided to avoid the complexities of the two fluid treatment (liquid-air) and preferred to adopt an *ad hoc* strategy: while the liquid in the vessel will be treated by standard SPH, the droplet will be treated as a set of points, whose interaction with themselves in the droplet will provide the surface tension, while the interaction with the liquid in the vessel is realized with another kind of force acting between each point and the SPH particle of the liquid in the vessel. Fig. 1 gives a sketch of the proposed interactions treatment.

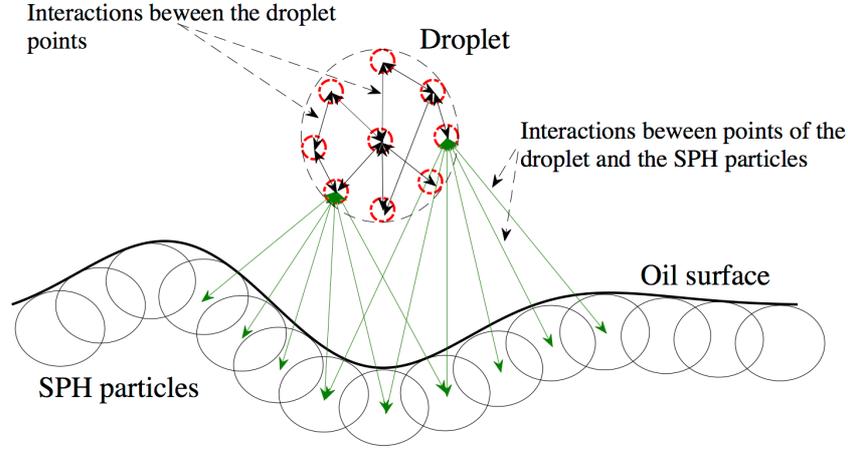

**Fig. 1:** schematic view of the interactions among the droplet points and between them and the SPH liquid particles. The interactions between the SPH particles are not shown.

The "small group" model attributes to a single point, constituting a droplet component, an elastic force acting on all other droplet points. It allows droplet deformation and produces a fictitious tension, whose value, in this study, is not compared with the real one.

The force between point $i$ and point $j$, we name $\vec{T}_{i,j}$, is given by the following expression:

$$\vec{T}_{i,j} = -\frac{f}{n_{drop}^2}\left(\frac{r_{i,j} - \delta_{drop}}{\delta_{drop}}\right)\frac{\vec{r}_{i,j}}{r_{i,j}} \qquad (1)$$

where $r_{ij}$ is the distance between two points, i and j, in the droplet, $\delta_{drop}$ is the size of the spatial distribution of the initial regular array of the points in the droplet, $n_{drop}$ is the total number of points in the droplet, the intensity factor $f$ is given by $f = 0.1 c_0^2 \rho_0 2s$, where $c_0$ is the reference sound speed, $\rho_0$ is the reference density of the liquid, $s$ is the scaling size of the repulsive droplet-liquid force described subsequently.

2.1.2 The interaction force between droplet and liquid

The interaction between the droplet points (both single or group) and the SPH particles of the liquid in the tank is given by a repulsive force function, which should have the role of the air pressure.

Therefore, it should increase as the droplet-liquid distance decreases. We tested a variety of functions and find that these three produce good results

$$a.\ \vec{F}_{i,j} = F_0 e^{-\frac{r_{i,j}}{s}}\left(\frac{\vec{r}_{i,j}}{r_{i,j}}\right), \quad b.\ \vec{F}_{i,j} = F_0 e^{-\left(\frac{r_{i,j}}{s}\right)^2}\left(\frac{\vec{r}_{i,j}}{r_{i,j}}\right), \quad c.\ \vec{F}_{ij} = \begin{cases} F_0 \left(\frac{4s - r_{i,j}}{4s}\right)^2 \left(\frac{\vec{r}_{i,j}}{r_{i,j}}\right) \\ 0 \ for\ r_{i,j} \geq 4s \end{cases} \quad (2)$$

The factor $F_0$ is given by the formula $F_0 = \frac{2s}{n_{drop}} c_0^2 \rho_0$ in the case of droplet simulated with a group of points or by the formula $F_0 = 2s c_0^2 \rho_0$ for a droplet made by a single point. The typical scale dimension of the range of the force is given by $s$, the scaling size, to be chosen freely, but of the order of magnitude of the initial SPH particles separation $\delta$. The $1/n_{drop}$ term is introduced as a normalizing factor so that the force depends only on the size of the droplet and does not depend on the number of points composing the droplet.

This force is exerted between each point composing the droplet and all nearby liquid SPH particles. It goes to zero if the distance $r_{i,j}$ is larger than the liquid particle separation $s\sim\delta$. This force is also subject to a further condition to simulate the coalescence effect: if the droplet interacts with a number of liquid particles for a time interval greater than the stability time [13], it is switched off definitely to zero, producing the disappearance of the droplet, merging it into the liquid and stopping the simulation.

2.2 Equations of motion of the fluid

With our model, which avoids the two-phase air-oil fluid treatment, the fluid governing equations, written in a Lagrangian framework give for the mass conservation:

$$\frac{d\rho}{dt} = -\rho \vec{\nabla} \cdot \vec{u} \quad (3)$$

the momentum equation:

$$\frac{d\vec{u}}{dt} = -\frac{1}{\rho}\vec{\nabla}P + \vec{g}Q(t) + \vec{A} \qquad (4)$$

where $\frac{d}{dt} = \frac{\partial}{\partial t} + \vec{u} \cdot \vec{\nabla}$ is the total derivative, $\vec{u}$ is the fluid velocity, $P$ the fluid pressure, $\rho$ the density, $\vec{A}$ is the local acceleration (i.e. of the fluid elements at their positions) due to the interaction between the droplet and the fluid, evaluated below; $\vec{g}$ the gravitational acceleration. Since we choose to work in the reference system of the container, then both the gravity and the inertial acceleration act on the liquid and on the droplet. Therefore, the gravitational acceleration is multiplied by the factor $Q(t) = 1 - \xi \sin\left(2\pi \frac{t}{P}\right)$, where $\xi$ is the adimensional amplitude of the vibration acceleration of the vessel. The maximum amplitude of the vibration is $Z_{max} = \frac{\xi P^2}{4\pi^2}$, obtained integrating $Q(t)$ over time and assuming that the initial position and velocity of the vessel are at Z=0.

Each droplet point moves according to the equation

$$\frac{d\vec{u}_i}{dt} = (\vec{g}Q)_i - \frac{\vec{F}_{tot_i}}{m_i} \qquad (5)$$

where $m_i$ is the mass of the point and

$$\vec{F}_{tot_i} = \sum_{j=1}^{n_{drop}} \vec{T}_{i,j} + \sum_{l=1}^{n_{sph}} \vec{F}_{i,l} \qquad (6)$$

Note that the index $i$ refers to any point of the droplet.

$\vec{T}_{i,j}$ is given by Eq. 1 and $\vec{F}_{i,l}$ is given by the Eq. 2b. In this approach, the motion of the centre of mass of the droplet is not computed by a further analytical equation, but it is obtained as a consequence of the motion of points forming the droplet. $\vec{F}_{tot_i}$ is the total force on the point $i$ of the droplet. This force mimics the whole behaviour of air and tension forces. To conserve the momentum, the SPH liquid particles, that interact with the droplet points, are subjected to opposite "air" forces, so we add on each "$k$" SPH particle, also the reaction force $-\sum_{i=1}^{n_{drop}} \vec{F}_{i,k}$. Consequently the $\vec{A}$ term of Eq. 4 is given by $\vec{A}_k = -\frac{\sum_{i=1}^{n_{drop}} \vec{F}_{i,k}}{m_k}$. When the droplet is made by a single point there is no tension and internal structure. With this approach, the droplet can be large with a small number of points. The computational cost for the droplet-liquid interaction is proportional to $n_{drop} \cdot N_{SPH}$ where $N_{SPH} = \frac{L_X L_Z}{\delta^2}$, $L_X$ and $L_Z$ are the dimension of the tank. To reduce the computational cost, the interaction between points and SPH particles of the liquid in the vessel can be limited to the liquid surface SPH particles, that is nearly $\frac{L_X}{\delta}$.

## 3. The Smoothed Particles Hydrodynamics numerical scheme

3.1 The approximated equation for the liquid motion

The liquid equations are approximated by using the SPH formalism [14], according to which a physical quantity *A* can be approximated by the discretization of an integral convolution

$$A(r) = \int A(r')\delta(r - r')dr' \cong \sum_l A(r'_l)W(r - r'_l)\Delta r'_l \qquad (7)$$

where $\delta(r - r')$ is the Dirac function and $\Delta r'_l = \frac{m_l}{\rho_l}$. $W(r - r'_l)$ is the interpolating function, named kernel, approximating the Dirac function. It has to be continuous and with continuous derivatives over a compact support.

There are many ways to evaluate functions and derivatives. We do not reproduce the mathematical procedure to obtain such formulae since that they are abundantly reported in many widespread papers [14, 15]. We just remind the basics.

In order to evaluate the derivatives, for instance, one can start from equivalent analytical expressions. Each way has its *pro* and *contra* [15]. The following one is very convenient since its antisymmetric structure allows the conservation of the linear and angular momentum of the discretized formulation [9]:

$$\vec{\nabla}P = \rho\left[\vec{\nabla}\left(\frac{P}{\rho}\right) + \frac{P}{\rho^2}\vec{\nabla}\rho\right] =$$
$$= \int \delta(r-r')\rho\left[\vec{\nabla}\left(\frac{P}{\rho}\right) + \frac{P}{\rho^2}\vec{\nabla}\rho\right]dr' \cong \int W(r-r'_i)\rho\left[\vec{\nabla}\left(\frac{P}{\rho}\right) + \frac{P}{\rho^2}\vec{\nabla}\rho\right]dr' \quad (8)$$

Evaluating these formulae in discrete terms it follows that the Eq. 4 is, therefore, approximated as:

$$\frac{d\vec{u}_i}{dt} = -\sum_{j=1}^{N} m_j \left(\frac{P_i}{\rho_i^2} + \frac{P_j}{\rho_j^2} + \prod_{ij}\right)\vec{\nabla}W_{ij} + \vec{g}Q(t) + \vec{A}_i \quad (9)$$

where $\prod_{ij} = -\nu\left(\frac{\vec{u}_{ij}\cdot\vec{r}_{ij}}{r_{ij}^2+\varepsilon h^2}\right)$ is the artificial viscosity. It is required in almost all basic SPH methods to stabilize the fluid and avoid numerical instabilities [14]. Here the subscripts *i, j*, identify the indexes of liquid particles, *m* their masses and $\vec{r}_{ij} = \vec{r}_i - \vec{r}_j$ their relative distances. Therefore $W_{ij} = W(\vec{r}_{ij})$, $\vec{u}_{ij} = \vec{u}_i - \vec{u}_j$, and $A_i$ is given in previous section. The quantity *h* is called '*smoothing length*' and controls the interaction size between the particles and the factor $\varepsilon = 0.01$ prevents a singularity in the artificial viscosity term, when $r_{ij}$ goes to zero. Finally, the viscosity $\nu_{ij}$ is defined by $\nu_{ij} = \frac{\alpha h \bar{c}_{ij}}{\bar{\rho}_{ij}}$ where $\alpha$ is a control parameter ranging from 0.01 to 0.1 and *c* sound speed. All upper signed variables are averaged between the liquid *i* and *j* particles. The liquid incompressibility is obtained by the weakly compressible approximation [14]. In this approach, the liquid is not truly uncompressible and its pressure is obtained using the Tait equation, $P = \frac{\rho_0 c_0^2}{7}\left[\left(\frac{\rho}{\rho_0}\right)^7 - 1\right]$ where the sound speed $c_0$ is chosen 20 times greater than the maximum fluid speed typical of the system, so in our cases $c_0 = 20\sqrt{gZ_{max}}$.

The continuity equation is cast into a discrete formalism as follows:

$$\frac{d\rho_i}{dt} = -\sum_j (\vec{u}_j - \vec{u}_i)\vec{\nabla}W_{ij}m_j \quad (10)$$

A new feature, compared to the standard Monaghan formulation, has been implemented in our scheme according to Molteni and Colagrossi [16]: we adopt the diffusion of the density of the fluid which makes the pressure profile of the solutions well behaved, continuous, and contributes to the general stability of the method.

We added to the continuity equation the term $\zeta\vec{\nabla}(hc\vec{\nabla}\rho)$, where $\zeta$ is an a-dimensional coefficient of the order of unity. This expression is discretized according to the SPH prescription:

$$\zeta hc_i \sum_j \psi_{ij} \vec{\nabla}_{ij} W_{ij} m_{ij} \qquad (11)$$

where $\psi_{ij} = 2\frac{(\vec{u}_i - \vec{u}_j)\cdot \vec{r}_{ij}}{r_{ij}^2 + \epsilon h^2}$ is due to Morris [17].

We use the Wendland interpolating kernel, which exhibits very good stability against the tensile instability [18].

The advancement in time is obtained by a second order accurate method described in Monaghan [14], which is a variant of the leapfrog scheme.

The viscosity of the liquid is simulated by exploiting the artificial viscosity typically used in basic SPH algorithms [14]. The artificial viscosity is mainly used to avoid instabilities at shocks and operates only in compression of volumes. We used it also in the expansion to produce a true viscous effect. The resulting kinematical viscosity is given by $\upsilon = \frac{1}{8}\alpha hc_0$. This approach is simple and has the computational benefit that there is no need of extra routines for viscous diffusion. The only thing one has to take into account is that viscosity depends on the three parameters: sound speed, spatial resolution and artificial viscosity coefficient.

3.2 Boundary conditions

To confine the liquid in the container we tested two type of boundary conditions: mirror and fixed particles. In the first one we have a standard mirror particle scheme [14], in the second one a simple strata of fixed particles. In the case of mirror model, the friction due to the viscosity between the liquid and the boundaries is small, while in the fixed particles one, the energy dissipation at the boundaries is a bit higher so that, to produce the same effect obtained with mirroring technique, it is required a longer time or larger value of the vertical acceleration.

In real experiments the action of the boundaries is very relevant. The dynamics of the liquid in the layer between the liquid and the container's wall is not exactly resolved in our simulations and this is a further fact that inhibits a close comparison with experiments.

**4. Parameters**

4.1 Some typical values

We tested many cases: different tank sizes both in $Z$ level and in $X$ length. The length of the vessel ranges from 0.05 m to 0.2 m, the $Z$ liquid level ranges from 0.003 to 0.005 m. The SPH particle size ranges from $h$=0.0003 to 0.0005 m, and the particles initial separation $\delta$ from 0.0002 to 0.0004 m. The liquid particles are set up in equilateral triangles with side length equal to particles initial distance, forming a regular array. The typical number of neighbours around each particle ranges from 12 to 30, so the criteria for good evaluation of the derivatives are well satisfied. The resulting total number of particles ranges from a few thousands to 20,000 particles. To make the drop we added one extra particle above the liquid surface, or we generated a group of few points: 9, 12, 16, 20, 35, 40. We stress that the standard tension formulae, used in the SPH approach to simulate the surface tension [10], are unable to prevent fragmentation for such a small particle number, so we used in our model the new interaction force described previously to obtain the tension effect.

4.2 Simulation's results

We tested the following cases with increasing period of the vibration $P$ = 0.015, 0.020, 0.030, 0.050 s. Here we discuss in detail the low period case $P$ = 0.015 $s$, which better suits for a comparison with the experimental data.

### 4.2.1 The P=0.015 s case

We simulate a container of length $L_X$ = 0.1 m, $L_Z$ = 0.005 m, h = 0.0003 m, $\delta$ = 0.00024 m, we used the fixed particles boundary conditions and the following parameters: artificial viscosity $\alpha$ = 0.025 roughly corresponding to a real viscosity $v = 8.46 \ 10^{-6} m^2/s$

Table 1 shows the behaviour of the finite size drops under different amplitudes of the tank oscillation; all other parameters are constant: tank dimensions, initial conditions, viscosity of the liquid, repulsive function type Gaussian, repulsion scale s = 0.00025 m. In the state *W* the droplet clearly has a walking behaviour, state *WNW* means that there are discontinuous episodes of walking periods (the drop alternates times in which it is stopped at times in which it moves with a constant speed, and the larger is the amplitude of oscillation the smaller becomes the inactivity periods), in the state *DW* it walks but the Faraday instability is strongly disturbing the drop motion, state *NW* means that it clearly doesn't walk. So, for instance, a drop made by 12 particles subjected to $\xi$ = 4 vibration amplitude walks in a very clean way.

| $\xi$ / $N_{drop}$ | 3 | 3.5 | 4 | 4.5 | 5 | 5.5 | 6 |
|---|---|---|---|---|---|---|---|
| 4 | WNW | WNW | WNW | WNW | WNW | WNW | DW |
| 6 | NW | W | W | W | W | W | DW |
| 9 | NW | W | W | W | W | W | DW |
| 12 | NW | NW | W | W | W | W | DW |
| 16 | NW | NW | W | W | W | W | DW |
| 20 | NW | NW | W | W | W | NW | DW |
| 35 | NW | NW | W | W(*) | W(*) | W(*) | W(*) |
| 40 | NW | NW | NW | NW | NW | NW | W(*) |

**Table 1**: Behaviour of the drops for P=0.015 s, (*) for short time intervals the drop has zero speed

Figs. 2, 3, 4 refer to the case vibration amplitude $\xi$ = 3.5 and reference sound speed 20 m/s. The drop is made by 9 points and the interaction force with the liquid is given by the Gaussian function. The total number of particles is $N = 5,861$. In this case, we have a "canonical" behaviour: wave and droplet travel at the same speed of about 0.06 m/s and the jumping period of the drop is equal to twice the exciting one. Fig. 2 shows the SPH particles, the droplet points and their speeds.

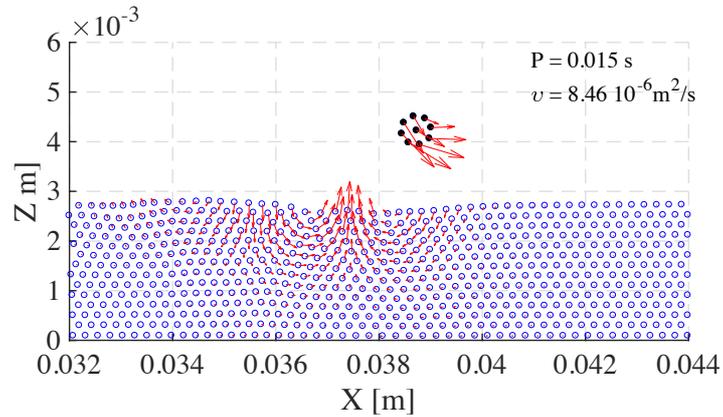

**Fig. 2**: Points composing the droplet and SPH particles of liquid bath with relative vectors velocity are shown. The horizontal extension is a portion of the tank.

The following colour figures show the liquid wave in a better way than the figure with the velocity vectors. It is worth noting that in our simulations the finite sized particle contra rotates, i.e. it rotates in the sense inverse to a wheel running on a surface. In our opinion, this effect may be due to the discrete nature of the simulation and to the asymmetric shape of the wave upon which the droplet falls, inducing a rotational motion. In this study, we did not focus on this effect since it seems irrelevant to the general droplet behaviour. The same zone with the vertical speed given by the colour scale is shown in Fig. 3.

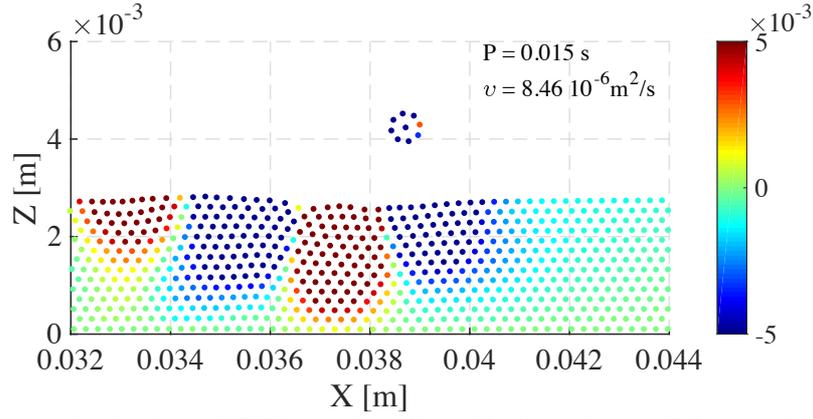

**Fig. 3**: Points composing the droplet and SPH particles of liquid bath are shown. Colours in the bar on the right-side change according with the value of the vertical velocity component $u_Z$. The rotating droplet exhibits coloured spots, accordingly. (For interpretation of the references to colour in this figure legend, the reader is referred to the web version of this article.).

Fig. 4 shows the same zone but with colours horizontal velocity component $u_x$.

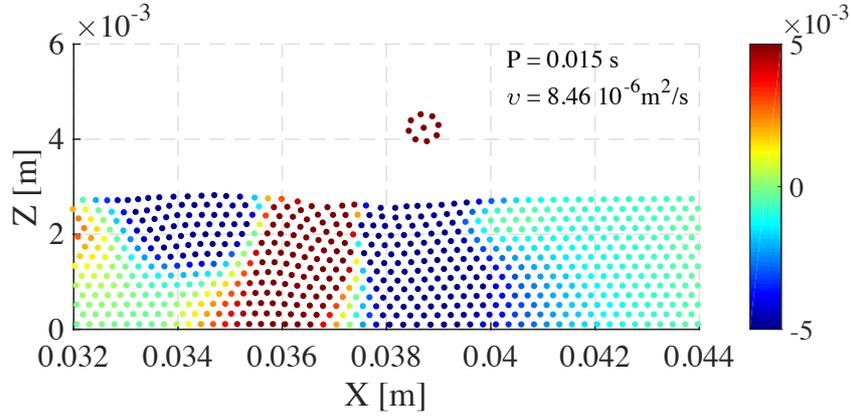

**Fig. 4:** Points composing the droplet and SPH particles of liquid bath are shown. Colours in the bar on the right-side change according with the value of the horizontal velocity component $u_X$. (For interpretation of the references to colour in this figure legend, the reader is referred to the web version of this article.)

Fig. 5 shows the minimum distance ($d_{min}$) of the drop from the liquid surface, the Z position and the scaled $u_x$, $u_z$ velocity component of the droplet versus time, the grid is set at $P = 0.015$ s intervals. The plot refers to a detailed view in a short temporal (0.5 - 0.7 s) window of the data produced by a simulation with amplitude of vibration $\xi = 3.30$. The speed is nearly constant and its profile is very similar to the experimental one as appearing in Fig. 2 of the paper by Milewski et al. [7]. It is clearly apparent the double period bouncing. This has a special value since Milewski behaviour occurs in a restricted, fine-tuned, range of the parameters values. The wiggle in the lines (especially in the one presenting the *Z* coordinate of centre of mass of the droplet) is due to the inertial forces since the physical quantities, here plotted, are computed in the reference frame moving with the container.

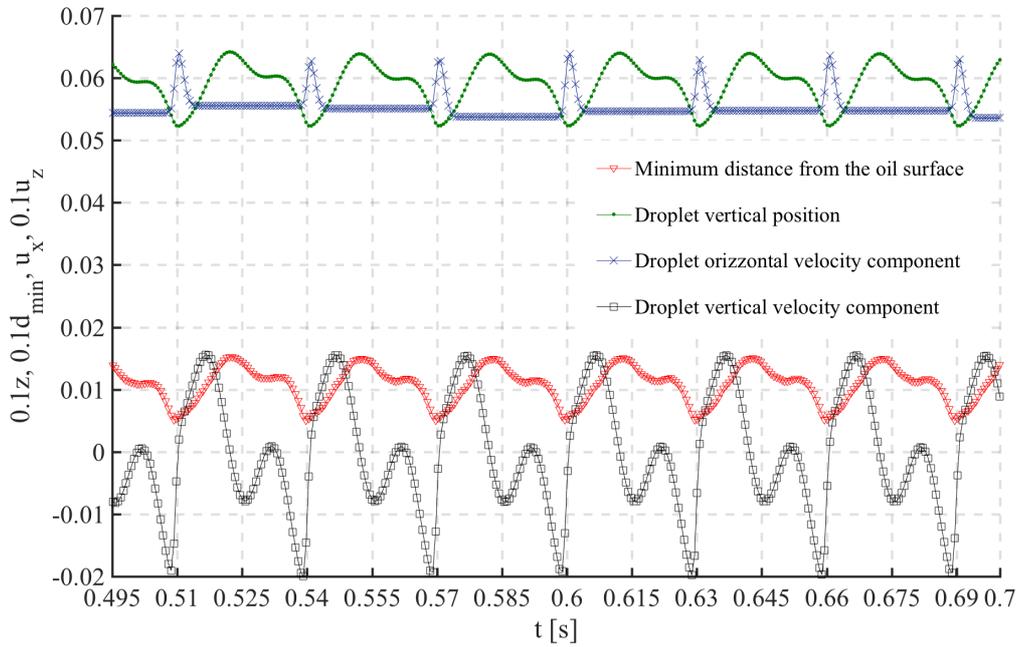

**Fig. 5:** Graph reports the minimum distance ($d_{min}$) of the droplet from the liquid surface, the *Z* position of the droplet centre of mass (divided by a factor 10), the $u_x$ (not scaled) and $u_z$ velocity component (divided by a factor 10) versus time.

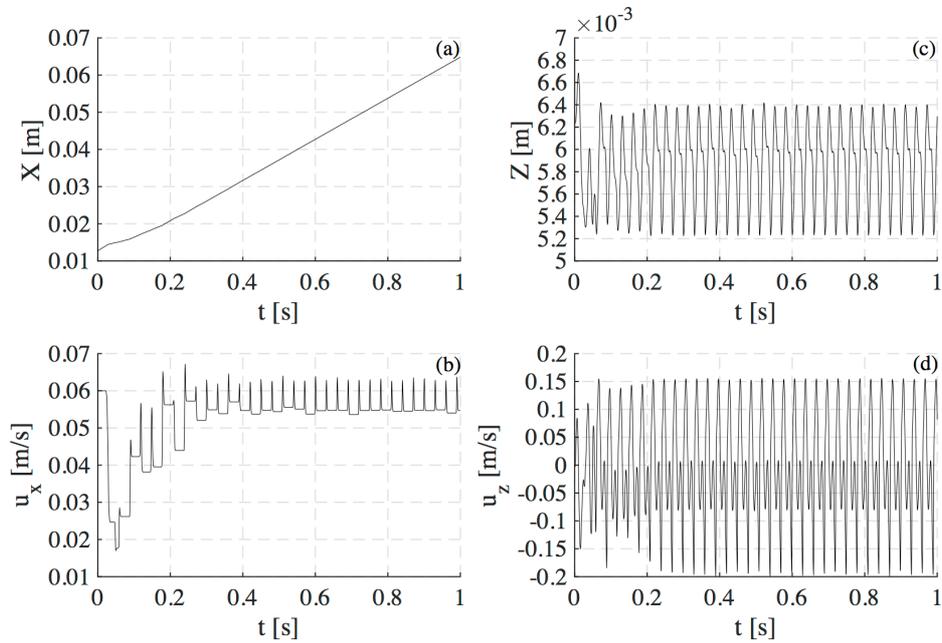

**Fig. 6:** (a) *X* and (c) *Z* coordinates of the centre of mass of the droplet, its velocity components (b) $u_x$ and (d) $u_z$, versus time.

In Fig. 6, *X* and *Z* coordinates of the centre of mass of the droplet, and its velocity components $u_x$ and $u_z$ versus time are shown. This is the case $\xi = 3.30$. The average velocity is about 0.056 m/s.

Fig. 7 shows the average velocity component $u_x$ of the droplet composed by 9 points versus different values of the vertical amplitude acceleration $\xi$. In the range of $\xi = 3.275 \div 3.650$ the droplet walks with an almost constant velocity, while, outside that range, the droplet still can walk but its speed suffers great variations around the average, from one jump to another. Below $\xi = 3.25$ the droplet does not walk.

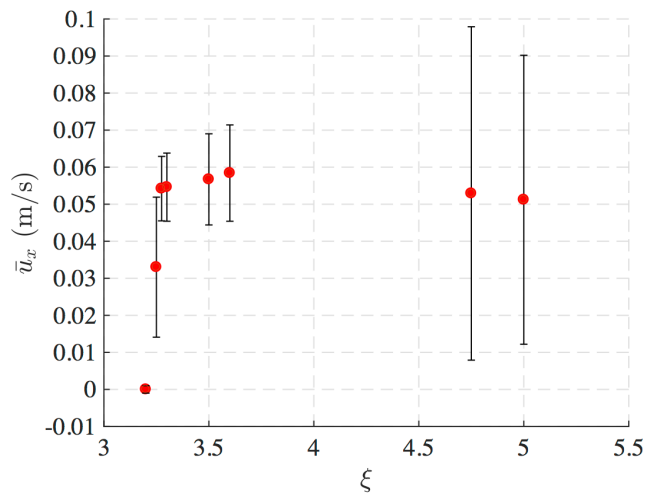

**Fig. 7**: Average velocity $u_x$ of the droplet made by 9 points versus amplitude of vertical acceleration.

4.2.2 The unstructured droplet cases

As already stated above, we also performed simulations of cases in which the droplet is made by a single point interacting with the liquid by the *ad hoc* forces described in previous section. We found that, for appropriate parameters, we obtained the walking behaviour. In our opinion, this fact is noticeable since it implies that the structure of the droplet is not critical; indeed, in these cases no surface tension is present. The droplet itself doesn't have a well definite size since it is the scale size of the repulsive force that gives an order of magnitude of the droplet size. Therefore, the mass of the droplet is a free parameter and its value is important to obtain the walking behaviour.

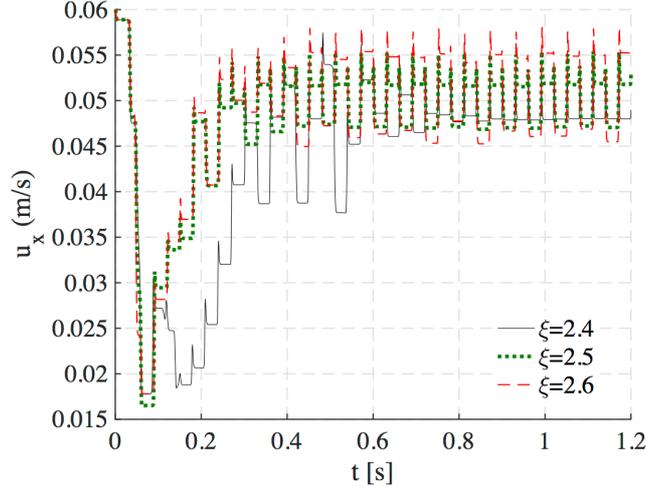

**Fig. 8:** $u_x$ component velocity of a point-like droplet for different amplitudes of the oscillation of the container.

In the Fig. 8 we show the $u_x$ component of the single point droplet corresponding to three different amplitudes of the sinusoidal oscillation of the container. The parameters of this simulation are: $P = 0.015$ s, length of the container $L_X = 0.1$ m, container's height $L_Z = 0.005$ m, SPH kernel size $h=0.0003$ m, initial particles separation $\delta = 0.00024$ m, sound factor 20; the repulsion function is Gaussian with scale size $s = 0.0005$ m. The boundaries are of "fixed particle" type, the total number of particles 11,956. The average velocity $u_x$ is the same, but the range of it variation increases with the amplitude of the forcing oscillation. Also in this case, the velocity profiles are similar to the ones shown in the paper by Milewski et al. [7], see in particular the case with amplitude of oscillation $\xi = 2.4$. The cases of $\xi = 2.5$ and $\xi = 2.6$ show regular two-level speed values.

Fig. 9 shows the average velocity $u_x$ and the difference between the maximum and minimum values of $u_x$ versus the amplitude of the oscillation of the vessel. It is quite clear that there is a sharp limit below which the walking phenomenon disappears and that as the amplitude values of the oscillation goes close to the Faraday instability the velocity component $u_x$ suffers stronger variations.

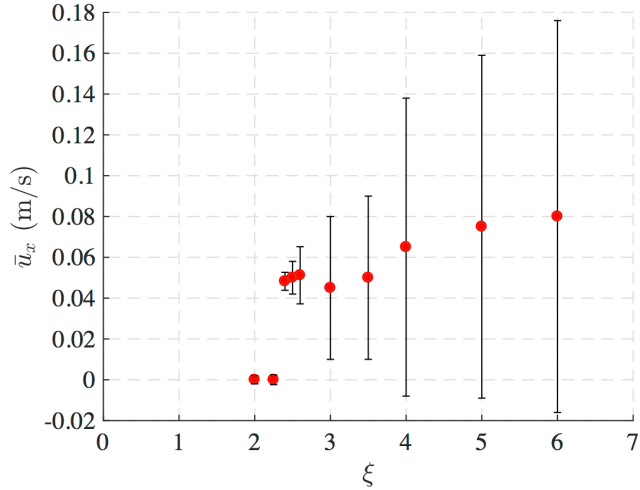

**Fig. 9:** Average velocity component u$_x$ versus vibration amplitude of the unstructured droplet.

Fig. 10 shows the *X-Z* positions of the SPH particles together with the unstructured droplet in a restricted zone of the vessel. The SPH particles and droplet have colours given by the velocity $u_z$, for the case with oscillation accelerating amplitude 2.6 times the gravitational one.

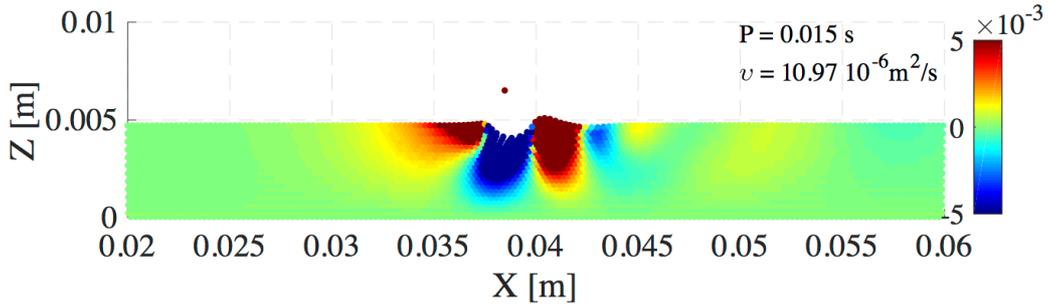

**Fig. 10:** *X-Z* view of the simulation results. Picture reports liquid particles in the tank and the point-like droplet. Colours in the bar on the right-side change according with the value of the vertical velocity component $u_z$.

## 5. Conclusions

We find that with an SPH Lagrangian code supplemented by ad hoc treatment of the droplet-liquid interaction, it is possible to obtain the basic walking droplet effect. In our study, the behaviour of our droplet is *similar* to the experimental one. In our simulations, when the wave-particle coupling is well established, the particle has an oscillating horizontal velocity whose average is constant. In general, this velocity depends on the amplitude of the vibration, the viscosity of the liquid and droplet size.

Here we show that, only in a restricted zone of the parameter space, the velocity is proportional to the amplitude of vibration.

We stress that also in the case of an unstructured, point-like droplet we obtain walking behaviour. In this case, the average velocity $u_x$ is weakly dependent from the amplitude of the vibrations, but its variation (i.e. maximum - minimum value) do increases with the amplitude of oscillations.

We emphasize the ineffectiveness of trying to reproduce an exact remake of the experiment unless a real full 3D simulation is set up. The real experiments are intrinsically 3D, so we remind that, even in an experimental situation where the tank is very thin in one dimension and very elongated in the other dimension, the interaction of the droplet with the air and liquid surface is still 3D.

We note also that the liquid motion in our *X-Z* simulations has clear vortices, so that we predict that a shallow wave approach, with its intrinsic average over the vertical direction, may suffer strong limitation to reproduce the phenomenon of the walking droplet.

It is clear to us that we are not exactly reproducing the Couder phenomenon, but we are producing a wave particle coupled object. This is the main result.

This study is provisional, however we preferred to publish these results to give colleagues a boost to study the phenomenon by numerical simulations.